\documentclass[twocolumn,aps,showpacs,amsmath,amsfonts,amssymb,floatfix,pre]{revtex4}
\usepackage{graphicx}
\usepackage{dcolumn}
\usepackage{bm}

\newcommand{\be}{\begin{equation}}
\newcommand{\ee}{\end{equation}}
\newcommand{\beq}{\begin{eqnarray}}
\newcommand{\eeq}{\end{eqnarray}}

\begin{document}
 
\title{Analytical results for stochastically growing networks: connection to the zero range process}
\author{P. K. Mohanty$^1$} \email{pk.mohanty@saha.ac.in}
\author{Sarika Jalan$^2$} \email{sarika@pks.mpg.de}
\affiliation{$^1$TCMP Division, Saha Institute of Nuclear Physics,1/AF,
Saltlake, Kolkata 700064, India\\
$^2$Max Planck Institute for the Physics of Complex Systems,
N\"{o}thnitzer Strasse 38, 01187 Dresden, Germany }
  
\begin{abstract}   

We introduce a stochastic model of growing networks where both, the number of new nodes 
which joins the network and the number of connections, vary stochastically.  We provide 
an exact mapping between this model and zero range process, and use this mapping to 
derive an analytical solution of degree distribution for any given evolution rule. One 
can also use this mapping to infer about a possible evolution rule for a given network. 
We demonstrate this for protein-protein interaction (PPI) network for Saccharomyces 
Cerevisiae.

\end{abstract}

\maketitle

Study of networks has been gaining recognition as a fundamental tool in understanding 
the dynamical behavior and response of real systems coming from different field such as 
biology, social systems, technological systems etc 
\cite{rev-Strogatz,rev-Barabasi,Newman,Amaral,rev-Costa}. Different network models have 
been proposed to study and understand these systems having underlying network structure. 
Erd\"{o}s and R\'{e}nyi random networks model was one of the oldest one, which shows 
that the probability ($p(k)$) of a node having degree $k$ follows exponential 
distributions, $p(k) \propto \exp(-k)$ \cite{ER}. Many real world networks however show 
scale-free behavior, $p(k)\propto k^{-\gamma}$, with the most striking examples of World 
Wide Web and cellular networks \cite{SF-www,SF-cellular} (for a review of scale-free 
networks refer \cite{rev-Barabasi}). In WWW, the number of incoming links follows
 power law with the value of $\gamma \sim 1.94$ \cite{SF-www} and analysis 
of metabolic networks of 43 organisms reveal that the number of chemical reactions (link) in
which a substrate (node) is involved in, show power law distribution, with
the exponent varying between 2.0 and 2.4 \cite{SF-cellular}.  

To capture scale-free behavior of real world networks, Barab\'{a}si-Albert (BA) proposed 
a growing network model based on the preferential attachment of the nodes 
\cite{BA,rev-Barabasi}. In the BA model each new node is connected with some old nodes 
with a probability {\it linearly} proportional to the degree of the node, $ u(k) \propto 
(k+\beta)$. This model gives rise to the scale-free network with degree distribution 
following power law $p(k) \propto k^{-\gamma}$, value of $\gamma=3+\beta$ \cite{BA}. 
Since then, several variations of BA algorithm have been proposed. An algorithm 
suggested by Dorgovtsev and Mendes based on the aging of the nodes also gives rise to a 
scale-free behavior \cite{Doro-aging}. Krapivsky et. al. also attempted to provide an 
analytical solution for different attachment function $u(k) \sim 
k^\lambda$\cite{Redner-GN}.

BA algorithm concentrates only on the degree distribution. Watts and Strogatz \cite{SW} 
proposed a model which captures the small diameter and large clustering properties shown 
by real world networks. Clustering coefficient basically measures the number of 
triangles, i.e. complete subgraphs or cliques of order $3$, in the network. Apart from 
cliques of the size 3, real world networks exhibit modular structures of higher levels 
\cite{modular}. For examples, in protein binding network of yeast \cite{cliques-pp} 
cliques of the size upto 14 nodes are present in the number much higher than 'random' 
\cite{cliques-random}. These small subgraphs are often considered to be building blocks 
of a network. Densities of a particular subgraph may tell if a network belongs to a 
certain superfamily \cite{superfamily} or perform specific functions \cite{Alon}. With 
all these insight into real world networks and in oder to capture these properties, 
particularly degree distribution and modules or cliques statistics, different other 
models \cite{model-Martinez,model-Redner} and evolution rules have been proposed 
\cite{evrule-Krapivsky}. In particular, Rozenfled and ben-Avraham \cite{{Daniel04}} 
proposed a local strategy for constructing scale-free network with external parameters 
capturing statistical properties of certain modular structures along with degree 
distribution.

In this paper we introduce stochasticity to the growing network models. Starting from 
the few initially connected nodes, a network in our model evolves as follows. At each 
time step, $n$ new nodes joins the network and make $m$ connections with existing nodes. 
Both $m$ and $n$ are taken as stochastic variables. Each new connection is made with a 
probability which depends on the degree of the node to be connected, need not be 
preferential. A special case of our model with linear connection probability and $n=1$, 
corresponds to the BA algorithm.  Note that our evolution rule, being stochastic, 
naturally captures various stochastic effects which are always present during the 
evolution of any real system.
 
First we show an explicit mapping between our model and the zero range process (ZRP), an 
exactly solvable model in non-equilibrium physics \cite{ZRP}, which provides an exact 
relation between any attachment rule $u(k)$ and the degree distribution $p(k)$ of the 
growing networks. So far there are several attempts to solve Barab\'{a}si-Albert model 
where $u(k)$ is linear in $k$, Dorgovtsev et. al. being the most close one 
\cite{Doro-BA}. These authors also did analytical calculations for certain other forms 
of preferential attachments \cite{Doro-PRE2001}. Krapivsky et. al. \cite{Redner-GN} have 
given analytical solution for $u(k) \sim k^\lambda$. Here, we provide exact degree 
distribution for any arbitrary evolution rule $u(k)$. This relation, being exact, can be 
inverted to 
infer about a possible evolution rule for any given real-world network.  Second, we show 
that the choice of stochastic parameters do not alter the degree distribution of the 
network. It only affects the correlations or statistical properties of the modules. 
Lastly we apply our methodology to a real world network and derive an stochastic 
evolution rule which captures the exact degree distribution. We argue that this method 
can be used to generate a growing network with any desired degree distribution.

First , the model. A generic algorithm for a growing network would be as follows. 
Starting from a small connected network, say with two nodes which are connected by a 
link, one brings $n$ new nodes at each iteration time $t$ and then each of these $n$ 
nodes connects to $m(i),i=1,n$ existing nodes. In general, $n$ and $m$ are 
stochastically varying positive integers drawn from distributions $\eta(n)$ and $h(m)$ 
respectively. These variations are not just the generalizations of \cite{BA}, it is 
quite natural that at some time variable number of nodes join realistic networks and 
make connections which vary from one node to the other. The probability that any given 
new node $i$ makes a link with one of the existing node $j$ is $w(k(j),t)$, where $k(j)$ 
is the degree of $j$ and $\sum_j w(k(j),t)=1$.

Now, let us find the steady state degree distribution $p(k)$ of these generic networks 
as $t\to \infty$. Let $M(k,t)$ be the number of nodes having $k$ links at time $t$.  
Since $\sum_k w(k,t)=1$, we may take $w(k,t)= u(k)/v(t)$ where
\be 
v(t) = \sum_k u(k) M(k,t).
\ee 
Here $u(k)$ is considered to be a generic function, need not be an increasing function 
which corresponds to the preferential attachment \cite{BA,Redner-GN}. The rate of 
increase of $M(k,t)$ is, then, given by
\beq
\frac{dM(k,t)}{dt} &=& \bar m \bar n\left[ \frac{u(k-1)}{v(t)}  M(k-1,t) - \frac{u(k)}{v(t)}  M(k,t) \right]\cr &&~~~~~~~~~~~~~+ \bar n h(k)
\label{eq_Mdot} 
\eeq
where $\bar n=\sum n \eta(n) $ is the average number of nodes which joins the network in 
each iteration step $t$. Equation (\ref{eq_Mdot}) is constrained by by $M(0,t)=0$, which 
ensures that every node in the network has nonzero links. The initial condition is 
$M(k,0)= 2 \delta_{k,1}$, $i.e$, we start with two nodes which are connected. Of course 
(\ref{eq_Mdot}) must be supplemented by the equation of growth rate of nodes,
\begin{equation}
\frac{dN(t)}{dt} = \bar n . 
\label{eq_Ndot} 
\end{equation}
In general, $\bar n $ may explicitly depend on $t$ if $\eta$ explicitly depend on $t$. 
We will considered this case later in this article. The degree distribution $p(k)$ in 
the steady state is defined as,
\be
 p(k) = \lim_{t\to \infty} \langle \frac{M(k,t)}{N(t)} \rangle, 
\ee
where averaging $\langle \dots \rangle$ is done over realizations. Clearly the steady-state 
is reached
only if  $M(k,t)\propto N(t)$  for large $t$. Thus in the steady state, we have 
\be 
M(k,t)= p(k) N(t).
\label{eq_product}
\ee
Here, we make an {\it ansatz} that the product form (\ref{eq_product})  holds even 
for large, but finite $t$. We will provide evidences in favor of this ansatz later in 
this article.
 
Using Eq. (\ref{eq_product}) one can rewrite (\ref{eq_Mdot}) as 
\be
\frac{1}{\bar m} \frac{v(t)}{N(t)} = \frac{u(k-1)p(k-1)-u(k)p(k)}{p(k)-h(k)}. 
\ee
Clearly, only a constant function, say $\alpha$,  satisfies above equation 
and we have, 
\beq
p(k) &=& \frac{u(k-1)} {\alpha +u(k)} p(k-1) +\frac{\alpha h(k)}{\alpha +u(k)}  
\label{eq_f} \\
\alpha&=&\frac{1}{\bar m} \frac{v(t)}{N(t)} = \frac{1}{\bar m}\sum_k  u(k) p(k).
\label{eq_alpha}
\eeq

There are few things to note here. First, that $\bar n$ do not appear in these 
equations. Thus, one may fix it to any arbitrary value without changing the degree 
distribution. We would argue and show later in this article that these irrelevant (with 
respect to degree distribution) parameters may marginally affect the correlations in the 
network. Second, that $p(k)$ is in fact normalized, which can be proved by summing 
Eq.~(\ref{eq_f}) for all $k$.

Solution of the difference equation (\ref{eq_f}) with natural boundary 
condition $p(0)=0$ can be written in a compact form
\be
p(k) = \frac{\alpha}{u(k)} \sum_{m=1}^k  h(m)\prod_{j=m}^k  \frac{u(j)}{\alpha +u(j)}.
\label{eq_pk}
\ee
However, the main difficulty remains in finding $\alpha$, which has to be self- 
consistently determined by using (\ref{eq_f})-(\ref{eq_alpha}).

First, let us consider the well studied case where at each time step only one node 
having $m_0$ links joins network. Then $\bar n =1$ and $h(m)= \delta_{m,m_0}$. Thus only 
a single term $m= m_0$ in Eq.~(\ref{eq_pk}) survives under the sum, and we have $p(k)=0$ 
for $k<m_0$. For $k \ge m_0$,
\be
 p(k) = \frac{\alpha}{u(k)}  \prod_{j=m_0}^k  \frac{u(j)} {\alpha+u(j)}
\label{eq_zrp}
\ee

If we use BA- algorithm  with  preferential attachment rule $u(k)= k+\beta$,
the degree distribution becomes   
\be
 p(k)=  \alpha \frac{\Gamma(\alpha +\beta +m_0)}{ \Gamma(1+\alpha +\beta +k)} 
 \frac{\Gamma(\beta+k)}{\Gamma(\beta+m_0) },
\label{eq_BA}
\ee
which can be used further to obtain $\alpha= 2+\beta/m_0$ from (\ref{eq_alpha}). Clearly 
for the large values of $k$, $p(k) \sim k ^{-1-\alpha}$. Thus the linear attachment rule 
$u(k) = k+\beta$, generates a scale-free network with $\gamma = 3+ \beta/m_0$. In the 
original formulation of Barab\'asi-Albert \cite{BA}, $\beta$ was taken to be zero and 
thus $\gamma=3$.

In the following we discuss the mapping of our growing network model with the ZRP. 
Eq.~(\ref{eq_zrp}) 
gives an explicit connection between the two. In ZRP, particles hop between the sites of a 
lattice with rate $w(k)$ where $k$ is the the occupancy of the departure site. The 
steady-state distribution of particles $\pi(k)$ in this model can be calculated exactly 
as $\pi(k) = {\cal N} \prod_{j=1}^k w(k)^{-1}$, where ${\cal N}$ is a normalization 
constant. From (\ref{eq_zrp}) one can identify that $\pi(k) = p(k) u(k)$ and then 
(\ref{eq_alpha}) becomes a normalization condition for $\pi(k)$. Corresponding rate is then
\be
w(k) = 
\left\{
\begin{array}{ll}
1 + \frac{\alpha}{u(k)}  & ~\textrm{ for } ~k\ge m_0 \\
1 &   \textrm{ for } ~ k< m_0 .
\end{array}
\right.
\label{eq_ZRP}
\ee

Now asymptotic behavior of $\pi(k)$, and thus $p(k)$, may be obtained from the known 
results of ZRP\cite{ZRP}.  To explain the importance of this mapping, let us take the 
example $u(k)= k^\lambda$ considered in \cite{Redner-GN}. There are following three different 
possibilities. For $0<\lambda<1$, $\pi(k)$ is a stretched exponential and thus $p(k) 
\sim \exp(-\alpha k^{1-\lambda} /(1-\lambda)) k^{-\lambda}$. For $\lambda=1$ one gets 
$p(k) \sim k^{-(\alpha+1)}$. Again, for $\lambda>1$, $\pi(k)$ asymptotically reaches a 
constant and thus we have distribution $p(k)~k^{-\lambda}$.

One can also obtain the asymptotic behavior of $p(k)$ by taking the continuum limit 
$x=k/K$ where $K$ is the maximum possible links (an arbitrarily large number). The 
difference equation (\ref{eq_f}) becomes a differential equation $$ - \frac{1}{p(x)} 
\frac{d}{dx} p(x)u(x)=\alpha, $$ with boundary condition $p(x_0)= 
\frac{\alpha}{u(x_0)}$, where $x_0= m_0/K$. A formal solution is then,
\beq
p(x) &=& \frac{\alpha}{u(x_0)} \frac{1}{u(x)} \exp \left( -\alpha \int^x \frac{dx^\prime}{u(x^\prime)}  \right) \\
\alpha &=& \frac{1}{x_0} \int dk  u(x) p(x)  
\label{eq_integrate}
\eeq 
It is easy to check that the above equations provide correct asymptotic values for exactly 
solvable cases, $u(k) =k+\beta$ and $u(k)= k^\lambda$.

Let us emphasize at this point that, although writing a close form expression for $p(k)$ 
for generic $u(k)$ is difficult, asymptotic behavior can be obtained easily using 
(\ref{eq_zrp})  or (\ref{eq_integrate}). As far as exact derivation of $p(k)$ is 
concerned, one may numerically implement (\ref{eq_pk}) and (\ref{eq_alpha}); $i.e.$, by 
iterating (\ref{eq_pk}) and (\ref{eq_alpha}), and assuming an initial $\alpha$. In most 
cases, we observe that $\alpha$ converges rapidly (within $15$ iterations) to a 
constant.

It is important to note that Eq. (\ref{eq_pk}) can be inverted   to  get
\be
u(k) = \frac{1}{p(k)} \sum_{i=1}^k \left[ h(i)-p(i)\right]
\label{eq_uk}
\ee

\begin{figure}
\includegraphics[width=8 cm]{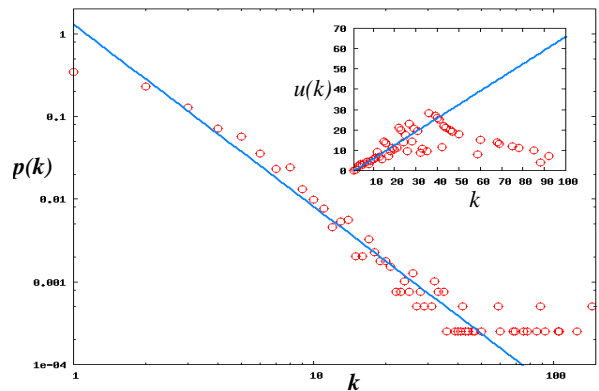}
\caption{Degree distribution for the PPI network for Saccharomyces Cerevisiae 
\cite{PPI-Yeast}. 
The evolution rule $u(k)$ derived using (\ref{eq_uk}) is shown in the inset. The solid line 
here (inset) is a linear fit $u(k)=k-.8$, for which one expects $p(k) \sim k^{-2.2}$. A 
solid line with slope $-2.2$ is drawn in the main figure to compare $p(k)$ with the theory.}
\label{fig_real}
\end{figure}

Here, $\alpha$ appears as an multiplicative constant which can be dropped as it is 
irrelevant for the evaluation of $p(k)$.  Eq.~(\ref{eq_uk}) provides an insight 
about a possible evolution rule for any real world network. For example we take 
protein-protein interaction (PPI) network for Saccharomyces Cerevisiae (yeast) 
\cite{PPI-Yeast}. The largest connected part has $N=3930$ nodes and 
$M = 7725$ links. The 
degree distribution of this network is shown in Fig.~\ref{fig_real}. The average degree 
of this network is $3.93$ which may be modeled using $h(m)=0.4 \delta_{m,1}+ 
0.234\delta_{m,2}+ 0.366\delta_{m,3}$. We evaluate $u(k)$ for this network (shown in the 
inset of Fig.\ref{fig_real}) using (\ref{eq_uk}) which fits well with a linear function 
$u(k)=1.5 (k-.8)$. Note that for this fitting we ignore large $k$ values as for these 
values, $p(k)$ is very small and sometimes $zero$ also. Corresponding degree 
distribution is now expected to be scale-free $p(k)\sim k^{-2.2}$, which is consistent 
with the observed distribution.

\begin{figure}
\includegraphics[width=6cm]{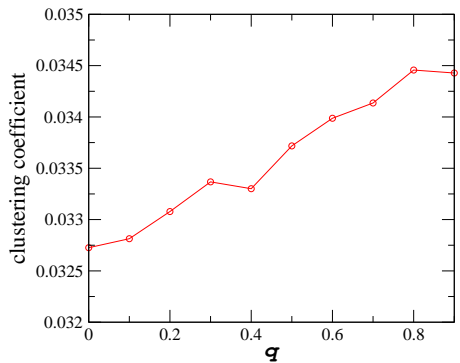}
\caption{Clustering coefficient changes with a stochastic parameter $q$ (see text). 
Other parameters are $u(k)=k+0.5$, $N=1000$, $m=4$, $\eta(n)= q\delta_{n,1} + (1-q) 
\delta_{n,5}$ and the clustering coefficient is averaged over $1000$ realizations.}
\label{qClus}
\end{figure}

Now, we turn our attention to the other stochastic parameters $\eta(n)$, namely the 
distribution of number of nodes which join the network during each iteration time step 
$t$. We have seen in (\ref{eq_pk}) that $\eta(n)$ do not alter the degree distribution. 
However they marginally affect correlations or the statistical properties of modular 
structures in the network. To illustrate this point, we generate a network with 
$u(k)=k+0.5$, $h(m)=\delta_{m,4}$ and $\eta(n)=q\delta_{n,1}+(1-q)\delta_{n,5}$, 
and measure the clustering coefficient for different $q$. As explained in the 
Fig.~(\ref{qClus}), we find that the clustering coefficient changes only marginally with 
$q$.

\begin{figure}
\includegraphics[width= 7 cm]{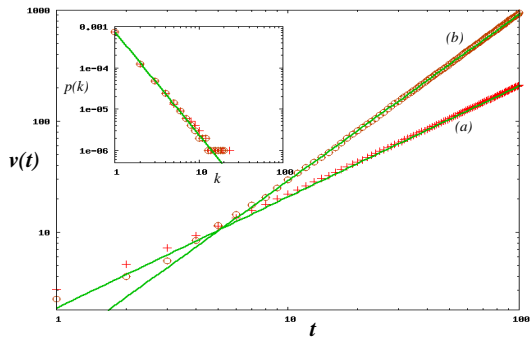}
\caption{Log scale plot of $v(t)$ for two different cases : (a) $\eta(n) = 
0.6\delta_{1,n} + 0.4\delta_{2,n}$ and (b) $n(t)= \sqrt t$. It is expected from 
(\ref{eq_Ndot}) and (\ref{eq_alpha}) that $v(t)$ is linear in first case, whereas for 
(b) $v(t) \sim t^{3/2}$. Solid lines with slope $1$ and $1.5$ are drawn for comparison.  
For both cases, $u(k) = k-.5$ and $m=1$ and averaging is done over $1000$ realizations. 
The degree distribution $p(k) \sim k^{-2.5}$ (inset) is identical for both cases.}
\label{fig_vt}
\end{figure}

Our analysis here rely on the fact that Eq.~(\ref{eq_product}) holds for large networks 
(as $t\to \infty$). Let us check the validity of (\ref{eq_product}) in details.  From 
(\ref{eq_alpha}) it is clear that $v(t)$ is proportional to $N(t)$ which can be 
obtained from (\ref{eq_Ndot}). First, we numerically evaluate $v(t)$ for few different 
networks and compare them with the theoretical results (\ref{eq_Ndot}). If the number of 
new nodes $n$ is a stochastic variable then $N(t)= \bar n t + 2$, is linear. However one 
can introduce an explicit time dependence in $n$ to get non-linear $N(t)$. For example, 
if $n(t)= \sqrt t$ we have $N(t) =t^{3/2}+2$ and thus $v(t) \propto t^{3/2}$. In figure 
(\ref{fig_vt}) we plot numerically measured $v(t)$ in log scale for two different cases; 
(a) $n = 0.6 \delta_{n,1} + 0.4 \delta_{n,2}$ and (b) $n(t) = \sqrt{t}$, both agree well 
with (\ref{eq_Ndot}). Although $N(t)$ is quite different, $p(k)$ (shown in the inset) 
was found to be same as expected. For both the cases evolution rule is $u(k)= k-.5$ and 
thus we have $p(k) \sim k^{-2.5}$.  To conclude, Eq.~(\ref{eq_product}) holds quite well 
after as few as ($t \sim 10$) iterations.  For large networks, the number of nodes which 
join in first few iteration steps is vanishingly small as compared to the size of the 
network, hence do not affect the network properties.

In summary, we introduce a generic model of stochastically growing network and show 
that this model can easily be mapped to the ZRP and thus enabling us to derive an exact 
relation between the degree distribution of network and its evolution function. This 
relation can be used to derive analytical form of the degree distribution for any 
arbitrary evolution rule and conversely for a given network data we can infer about a 
possible evolution rule. Our evolution rule produce exact degree distribution, as 
obtained from the given network data, even for small $k$ values. We demonstrate this by 
taking example of a real world PPI networks and deriving a possible evolution rule to this 
network.

Based on our exact calculations we expect to get the better understanding of the the 
evolution of real world networks. Also, since ZRP is exactly solvable, mapping of ZRP 
with network growth models, opens up a platform to study the interplay between evolution 
rules and steady state degree distribution.

One of us (PKM) acknowledges MPIPKS for the hospitality and the support.

\end{document}